# Bibliometric-enhanced Information Retrieval

Philipp Mayr*, Andrea Scharnhorst, Birger Larsen, Philipp Schaer, Peter Mutschke

* GESIS – Leibniz Institute for the Social Sciences, Unter Sachsenhausen 6-8, 50667 Cologne, Germany
`philipp.mayr@gesis.org`

**Abstract.** Bibliometric techniques are not yet widely used to enhance retrieval processes in digital libraries, although they offer value-added effects for users. In this workshop we will explore how statistical modelling of scholarship, such as Bradfordizing or network analysis of coauthorship network, can improve retrieval services for specific communities, as well as for large, cross-domain collections. This workshop aims to raise awareness of the missing link between information retrieval (IR) and bibliometrics/scientometrics and to create a common ground for the incorporation of bibliometric-enhanced services into retrieval at the digital library interface.

**Keywords:** Bibliometrics, Informetrics, Scientometrics, Information Retrieval, Digital Libraries

## 1    Background

The information retrieval (IR) and bibliometrics / scientometrics communities move more closely together with combined recent workshops like "Computational Scientometrics"[1] (held at iConference 2013 and CIKM 2013) and "Combining Bibliometrics and Information Retrieval"[2] (held in July at the ISSI conference 2013) which was organized by the authors of this workshop proposal. The ISSI workshop attracted more than 80 participants. The high interest among the bibliometricians was also generated by contributions from three "de Solla-Price"-medal winners and leading-edge bibliometricians Michel Zitt, Wolfgang Glänzel and Howard D. White. The main focus of their talks "Meso-level retrieval: field delineation and hybrid methods", "Bibliometrics-aided retrieval" and "Co-cited Author Maps, Bibliographic Retrievals, and a Viennese Author" was on the influences of IR on bibliometrics (e.g. as a tool to do better bibliometrics analyses). Two papers by Dietmar Wolfram and Birger Larsen highlighted the technical aspects of IR from a bibliometric viewpoint.

During these previous workshops it became obvious that there is a growing awareness that exploring links between bibliometric techniques and IR could be beneficial for actual both communities. They also made visible that substantial future work in this direction depends on a rise in awareness in both communities. IR and biblio-metrics go a long way back. Many pioneers in bibliometrics (e.g. Goffman, Brookes, Vickery), actually came from the field of IR, which is one of the traditional branches of information science. IR as a technique stays at the beginning of any scientometric exploration, and so, IR belongs to the portfolio of skills for any bibliometrician / scientometrician. However, IR and bibliometrics as special scientific fields have also grown apart over the last decades.

---

[1] http://www.cse.unt.edu/~ccaragea/CIKM-WS-13.htm
[2] http://www.gesis.org/en/events/conferences/issiworkshop2013

## 2   Goals, Objectives and Outcomes

Our workshop proposal aims to engage with the IR community about possible links to bibliometrics and complex network theory which also explores networks of scholarly communication. Bibliometric techniques are not yet widely used to enhance retrieval processes in digital libraries, yet they offer value-added effects for users (Mutschke et al., 2011). To give an example, recent approaches have shown the possibilities of alternative ranking methods based on citation analysis leading to an enhanced IR.

Our interests include information retrieval, information seeking, science modelling, network analysis, and digital libraries. The goal is to apply insights from bibliometrics, scientometrics, and informetrics to concrete practical problems of information retrieval and browsing.

Retrieval evaluations have shown that simple text-based retrieval methods scale up well but do not progress (Armstrong et al., 2009). Traditional retrieval has reached a high level in terms of measures like precision and recall, but scientists and scholars still face challenges present since the early days of digital libraries: mismatches between search terms and indexing terms, overload from result sets that are too large and complex, and the drawbacks of text-based relevance rankings. Therefore we will focus on statistical modelling and corresponding visualizations of the evolving science system. Such analyses have revealed not only the fundamental laws of Bradford and Lotka, but also network structures and dynamic mechanisms in scientific production (Börner et al., 2011). Statistical models of scholarly activities are increasingly used to evaluate specialties, to forecast and discover research trends, and to shape science policy (Scharnhorst et al., 2012). Their use as tools in navigating scientific information in public digital libraries is a promising but still relatively new development. We will explore how statistical modelling of scholarship (e.g. White et al., 2004) can improve retrieval services for specific communities, as well as for large, cross-domain collections. Some of these techniques are already used in working systems but not well integrated in larger scholarly IR environments.

The availability of new IR test collections that contain citation and bibliographic information like the iSearch collection (presented at the ISSI workshop by Birger Larsen, see Lykke et al., 2010) or the ACL collection (Ritchie, Teufel, and Robertson, 2006) could deliver enough ground to interest (again) the IR community in these kind of bibliographic systems. The long-term research goal is to develop and evaluate new approaches based on informetrics and bibliometrics. More specifically, we ask questions such as:

- How can we build scholarly information systems that explicitly use these approaches at the user-system interface?
- Are bibliometric-enhanced retrieval systems a value-added for scholarly work?
- How can models of science be interrelated with scholarly, task-oriented searching?
- And the other way around: Can insights from searching also improve the underlying statistical models themselves?

Although IR and scientometrics belong to one discipline, they are driven by different epistemic perspectives. In the past, experts from both sides have called for closer collaboration, but their encounters are rather ad-hoc. This workshop aims to raise awareness of the missing link between IR and bibliometrics/scientometrics and to create a common ground for the incorporation of bibliometric-enhanced services into retrieval at the digital library interface.

## 3    Format and Structure of the Workshop

The workshop will start with an inspirational keynote to kick-start thinking and discussion on the workshop topic, e.g. by one of the organizers of the ACM SIGKDD 2013 Cup, which used a large dataset from Microsoft Academic Search in an Author-Paper Identification Challenge[3]. This will be followed by paper presentations in a format found to be successful at EuroHCIR this year: Each paper is presented as a 10 minute lightning talk and discussed for 20 minutes in groups among the workshop participants followed by 1-minute pitches from each group on the main issues discussed and lessons learned. The workshop will conclude with a round-robin discussion of how to progress in enhancing IR with bibliometric methods.

## 4    Audience

The audiences (or clients) of IR and bibliometrics are different. Traditional IR serves individual information needs, and is – consequently – embedded in libraries, archives and collections alike. Scientometrics, and with it bibliometric techniques, has matured serving science policy.
We propose a half-day workshop that should bring together IR and DL researchers with an interest in bibliometric-enhanced approaches. Our interests include information retrieval, information seeking, science modelling, network analysis, and digital libraries. The goal is to apply insights from bibliometrics, scientometrics, and informetrics to concrete, practical problems of information retrieval and browsing.

The workshop is closely related to the workshop "Combining Bibliometrics and Information Retrieval" held at ISSI and tries to bring together contributions from core bibliometricians and core IR specialists but having selected those who already operate on the interface between scientometrics and IR.

---

[3]    https://www.kaggle.com/c/kdd-cup-2013-author-paper-identification-challenge

## 5  Output

After the ISSI 2013 workshop on "Combining Bibliometrics and Information Retrieval" the workshop organizers were invited to apply for a special issue in Scientometrics. Such a dissemination serves well to account for raised awareness and contributions from the bibliometrics side and written for the bibliometrics side. We aim with the proposed workshop for a similar dissemination strategy, but now oriented towards core-IR. This way, we build a sequence of explorations, visions, results documented in scholarly discourse, and set up enough material for a sustainable bridge between bibliometrics and IR.

### References

1. Armstrong, T. G., Moffat, A., Webber, W., & Zobel, J. (2009). Improvements that don't add up: ad-hoc retrieval results since 1998. In *Proceeding of the 18th ACM Conference on Information and Knowledge Management* (pp. 601–610). Hong Kong, China: ACM. doi:10.1145/1645953.1646031
2. Börner, K., Glänzel, W., Scharnhorst, A., & van den Besselaar. P. (2011). Modeling science: Studying the structure and dynamics of science." *Scientometrics* 89, 347–348.
3. Lykke, Marianne, Birger Larsen, Haakon Lund, and Peter Ingwersen. (2010). "Developing a Test Collection for the Evaluation of Integrated Search." In *Advances in Information Retrieval*, edited by Gurrin et al., 5993:627–630. Lecture Notes in Computer Science. Berlin, Heidelberg: Springer.
4. Mutschke, P., Mayr, P., Schaer, P., & Sure, Y. (2011). Science models as value-added services for scholarly information systems. *Scientometrics*, *89*(1), 349–364. doi:10.1007/s11192-011-0430-x
5. Ritchie, Anna, Simone Teufel, & Stephen Robertson. (2006). "Creating a Test Collection for Citation-based IR Experiments." In *Proceedings of the Main Conference on Human Language Technology Conference of the North American Chapter of the Association of Computational Linguistics*, 391–398. HLT-NAACL '06. Stroudsburg, PA, USA: Association for Computational Linguistics. doi:10.3115/1220835.1220885. http://dx.doi.org/10.3115/1220835.1220885.
6. Scharnhorst, A., Börner, K., & Besselaar, P. van den (Eds.). (2012). *Models of Science Dynamics Encounters Between Complexity Theory and Information Sciences*. Berlin: Springer.
7. White, H.D., Lin, X., Buzydlowski, J.W., & Chen, C. (2004). User-controlled mapping of significant literatures. *Proceedings of the National Academy of Sciences* 101 (suppl. 1), April 6, 2004, 5297-5302.

## Short bios of the proposers

**Philipp Mayr**

Philipp Mayr is a postdoctoral researcher and team leader at the GESIS – Leibniz Institute for the Social Sciences department Knowledge Technologies for the Social Sciences. Philipp is a graduate of the Berlin School of Library and Information Science at Humboldt University Berlin where he finished his doctoral research in 2009. Philipp is a member of the European NKOS network and published widely in the areas Informetrics, Information Retrieval and Digital Libraries. He is member of the editorial board of the journals Scientometrics and Information - Wissenschaft & Praxis. His research interests include non-textual ranking in digital libraries, bibliometric methods, evaluation of information systems and knowledge organising sytems, as well as applied informetrics http://www.ib.hu-berlin.de/~mayr/. Philipp was the main organizer of the workshop "Combining Bibliometrics and Information Retrieval" at ISSI 2013.

**Andrea Scharnhorst**

Dr. Andrea Scharnhorst is Head of e-Research at the Data Archiving and Networked Services (DANS) institution in the Netherlands - a large digital archive for research data primarily from the social sciences and humanities. She is also coordinates the computational humanities programme at the e-humanities group of the Royal Netherlands Academy of Arts and Sciences (KNAW) in Amsterdam. Starting in physics (Diploma in Statistical Physics) she got her PhD in philosophy of science. She co-edited books in the Springer Series of Understanding Synergetics on Innovation Networks (with A. Pyka) and recently on Models of Science Dynamics (with K. Boerner and P. van den Besselaar). Her current work in the information sciences is devoted to the development of *knowledge maps* for library collections, research data bases and on-line knowledge spaces such as Wikipedia. Andrea was co-organizer of the workshop "Combining Bibliometrics and Information Retrieval" at ISSI 2013.

**Birger Larsen**

Birger Larsen is (from October 1, 2013) professor of Information Analysis and Information Retrieval at the Department of Communication at Aalborg University, Copenhagen. His main research interests include Information Retrieval (IR), structured documents in IR, XML IR and user interaction, domain specific search, understanding user intents and exploiting context in IR, as well as Informetrics/Bibliometrics, citation analysis and quantitative research evaluation. He is part of the team behind the iSearch test collection, which with 450.000+ scientific documents is one of the largest available test collections that facilitate experiments with both scientific documents and citations networks. He is broadly engaged in program committees in the main journals and conferences within the areas covered by his research interests, and is a frequent involved as organiser of workshops, symposia and conferences, e.g. as PC co-chair of ISSI2009, general co-chair of ICTIR2013, co-chair of the EuroHCIR workshop series, co-chair of the ECIR2012 workshop on Task Based and Aggregated

Search, etc. Birger was the author of an invited paper at the workshop "Combining Bibliometrics and Information Retrieval" at ISSI 2013.

**Philipp Schaer**

Philipp Schaer is a postdoctoral researcher at GESIS – Leibniz Institute for the Social Sciences. He studied computer sciences at University of Koblenz where he received his master's degree and his Ph.D. The main part of his research is in the fields of digital libraries, information retrieval and the application of informetrics in the these fields. Philipp was co-organizer of the workshop "Combining Bibliometrics and Information Retrieval" at ISSI 2013.

**Peter Mutschke**

Peter Mutschke is senior researcher at GESIS – Leibniz Institute for the Social Sciences (Cologne) and acting head of the GESIS department "Knowledge Technologies for the Social Sciences". His research focuses on information retrieval, network analysis and Social Web. He worked in a number of national and international research projects such as DAFFODIL (Distributed Agents for User-Friendly Access of Digital Libraries ), INFOCONNEX (interdisciplinary information network for Social Sciences, Education Science and Psychology), IRM (value-added search services), the DELOS/NSF Working Group on reference models for digital libraries, and the EU-funded project WeGov (Where eGovernment meets the eSociety). Currently, he is involved in major national and European research networks such as the COST action "Analyzing the dynamics of information and knowledge landscapes" (KNOWeSCAPE) and the research network "Science 2.0" of the German Leibniz Association. For both research networks Peter Mutschke is member of the management committee. Peter Mutschke is author of a number of research articles, member of a number of international programme committees, and was co-organizer of the workshop "Combining Bibliometrics and Information Retrieval" at ISSI 2013.